\documentclass{article}

\usepackage{fullpage}
\usepackage{natbib}
\usepackage{amsmath,amsfonts,amsthm,amssymb}
\usepackage{graphicx}
\usepackage[ruled,linesnumbered]{algorithm2e}
\usepackage{tikz}
\usepackage{multirow} 
\usetikzlibrary{calc,positioning}
\usepackage{authblk}
\usepackage{ulem}

\title{Spatial modeling of extremes and an angular component 
}
\author{G. Tamagny}
\author{M. Ribatet}
\affil{Nantes Université, École Centrale Nantes, CNRS, Laboratoire de
  Mathématiques Jean Leray, LMJL, UMR 6629, F-44000 Nantes, France} 
\date{}
\begin{document}
\maketitle
\begin{abstract}
  Many environmental processes such as rainfall, wind or snowfall are
  inherently spatial and the modelling of extremes has to take into
  account that feature. In addition, environmental processes are often
  attached with an angle, e.g., wind speed and direction or extreme
  snowfall and time of occurrence in year. This article proposes a
  Bayesian hierarchical model with a conditional independence
  assumption that aims at modelling simultaneously spatial extremes and
  an angular component. The proposed model relies on the extreme value
  theory as well as recent developments for handling directional
  statistics over a continuous domain. Working within a Bayesian
  setting, a Gibbs sampler is introduced whose performances are
  analysed through a simulation study. The paper ends with an
  application on extreme wind speed in France. Results show that
  extreme wind events in France are mainly coming from West apart from
  the Mediterranean part of France and the Alps.
\end{abstract}

\noindent%

\section{Introduction}

The accurate modelling of environmental extremes such as floods or
heatwaves is crucial as they can lead to severe damages or economic
losses. As a consequence, the last decade has seen many theoretical
developments as well as extensive applications to try to analyse
spatial extremes. From a probabilistic point of view, the relevant
framework for modelling extreme events is the extreme value theory
whose focus is precisely on the tail of the distribution.

In a spatial setting, Bayesian hierarchical models or max-stable and
Pareto processes \citep{deHaan1984,deHaan2014,Dombry2015} are popular
choices. While the latter approaches are relevant for modelling areal
quantities as cumulative rainfall amount over a spatial domain,
pointwise predictions of extremes is still of interest in many
applications, e.g., design of dykes for floods. In this context,
Bayesian hierarchical models have been found to be a very competitive
choice and widely applicable as opposed to max-stable or Pareto
processes for which inference is challenging and pointwise prediction
less accurate \citep{Dombry2017,Huser2019,Davison2012}. As a
consequence and, as long as pointwise prediction is of concern, one
can restrict our attention to the univariate case for which extreme
value theory justifies the use of the Generalized Extreme Value
distribution (GEV) whose cumulative distribution function is
\begin{equation}
  \label{eq:GEV}
  \Pr\{\eta(s) \leq z\} = \exp\left[- \left\{1 + \xi(s) \frac{z - \mu(s)}{\sigma(s)}
    \right\}^{-1/\xi(s)} \right], \qquad \sigma(s) + \xi(s) (z - \mu(s)>
  0,
\end{equation}
where $\eta(s)$ denotes the block maxima random variable at location
$s \in \mathcal{X}$, and $\mu(s) \in \mathbb{R}$,
$\sigma(s) \in (0, \infty)$, $\xi(s) \in \mathbb{R}$ are respectively
the location, scale and shape GEV parameters.

For most applications, it is sensible to let the GEV parameters vary
over the spatial domain $\mathcal{X}$ so that one must impose a
regression structure on these parameters, e.g.,
$\mu(s) = f_\mu\{\mathbf{x}(s); \boldsymbol{\beta}_\mu\}$ where
$\mathbf{x}(s)$ denotes a vector of covariates evaluated at location
$s$. A popular choice is to assume a linear form, e.g.,
$f_\mu(\mathbf{x}(s); \boldsymbol{\beta}_\mu) = \mathbf{x}^\top(s)
\boldsymbol{\beta}_\mu$ but often yields to unrealistically smooth
surfaces \citep{Davison2012}. To bypass this hurdle, one can take
benefit of using a latent hierarchical model with a conditional
independence assumption \citep{Cooley,Casson1999} where the GEV
parameters now vary spatially according to some stochastic process,
i.e.,
\begin{align}
\label{eq:modelCooley}
    \eta(s) \mid \{\mu(s), \sigma(s), \xi(s)\}
  &\stackrel{\text{ind}}{\sim} \mbox{GEV}\{\mu(s), \sigma(s),
    \xi(s)\}, \qquad s \in \mathcal{X}  \\ 
  \notag
  \mu(\cdot) &\sim \mbox{GP}(m_\mu, \gamma_\mu)\\
  \notag
  \sigma(\cdot) &\sim \mbox{GP}(m_\sigma, \gamma_\sigma)\\
  \notag
  \xi(\cdot) &\sim \mbox{GP}(m_\xi, \gamma_\xi)
\end{align}
where $\mbox{GP}(m, \gamma)$ denotes a (univariate) Gaussian process
with mean function $m\colon s \mapsto m(s)$ and covariance function
$\gamma\colon (s_1,s_2) \mapsto \gamma(s_1,s_2)$ and the three
Gaussian processes above are usually assumed independent for
simplicity.

To the best of our knowledge, statistical developments have focused
mainly on the magnitude of extremes, e.g., rainfall amount, while
little attention has been paid to some additional features of these
events. More precisely, environmental processes are often attached
with an angular component, e.g., wind direction. One exception is
proposed by \citet{Konzen2021} but the aim was to handle appropriately
angular covariates and not any angular outcome. This paper aims at
filling this gap by considering a hierarchical Bayesian framework
which takes advantage of the conditional independence assumption and
is able to handle both extremal and angular features of the data. The
paper is organized as follows. Section~\ref{sec:proj-gauss-proc}
introduces the projected Gaussian process
model. Section~\ref{sec:extr-angul-bayes} gives an extension of
model~\eqref{eq:modelCooley} by adding an additional layer
corresponding to the projected Gaussian model and details how
inference can be done using MCMC
techniques. Section~\ref{sec:simulation-study} gives results of a
simulation study while Section~\ref{sec:application} applies the
proposed methodology to extreme wind speeds in France. The paper ends
with a discussion. Specific details on MCMC implementation are
deferred to the appendix.

\section{The projected Gaussian process model}
\label{sec:proj-gauss-proc}

Parametric statistical models for directional data are often based on
three main families: distribution on $[0,2\pi]$, wrapped distribution
on $\mathbb{R}$ and projected distributions \citep{Mardia}. This paper
focuses on projection methods and more specifically on projected
Gaussian distributions as extension to the spatial setting is
straightforward. The projected Gaussian model \citep{wang} is defined
as a random angle $\theta$ obtained via the projection of a bivariate
Gaussian vector $X$ onto the unit circle
$\mathbb{S}^1 = \{x \mathbb{R}^2\colon \|x\|_2 = 1\}$.

\begin{figure}
    \centering
    \includegraphics[width=0.49\textwidth]{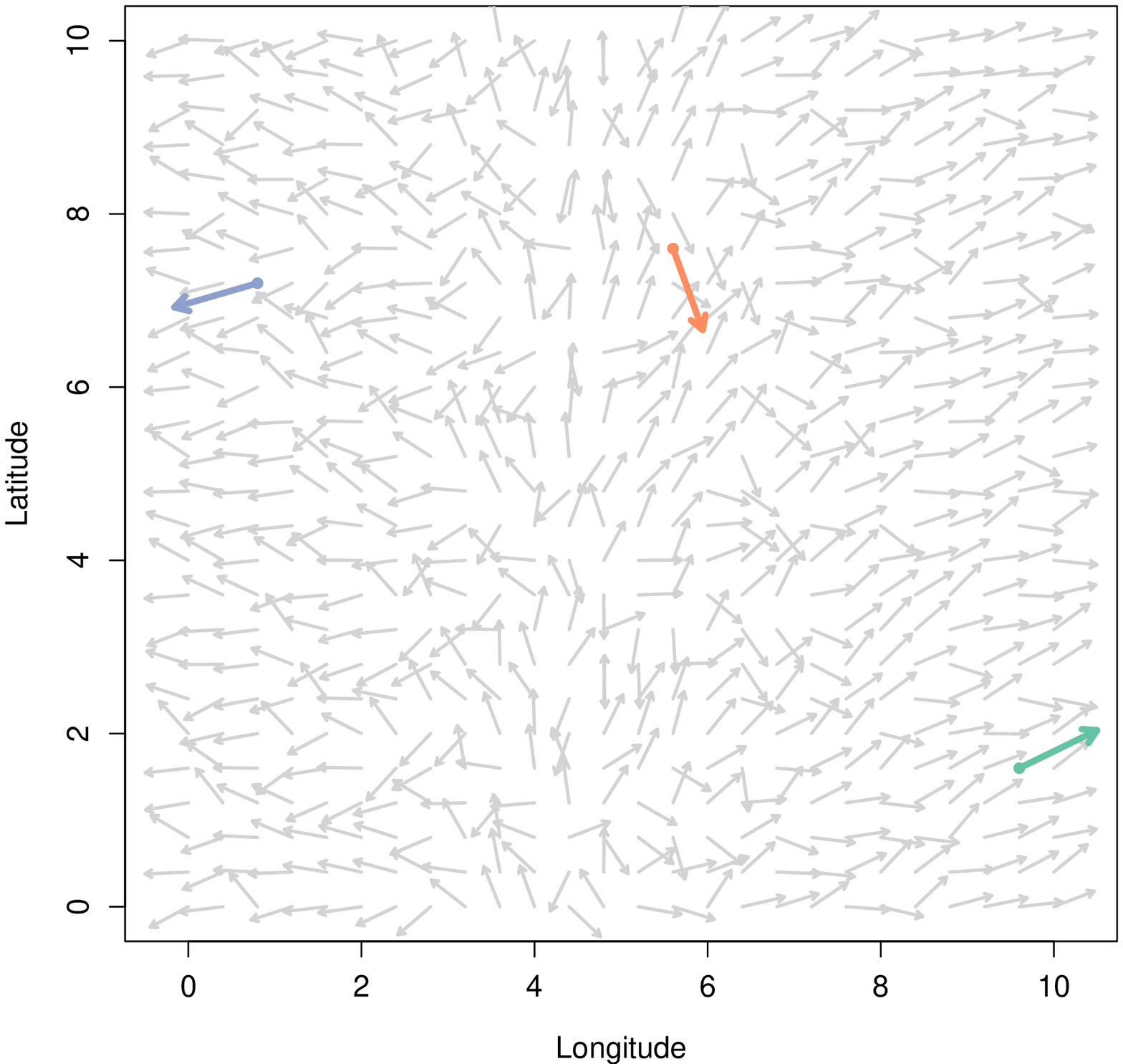}\hfill%
    \includegraphics[width=0.49\textwidth]{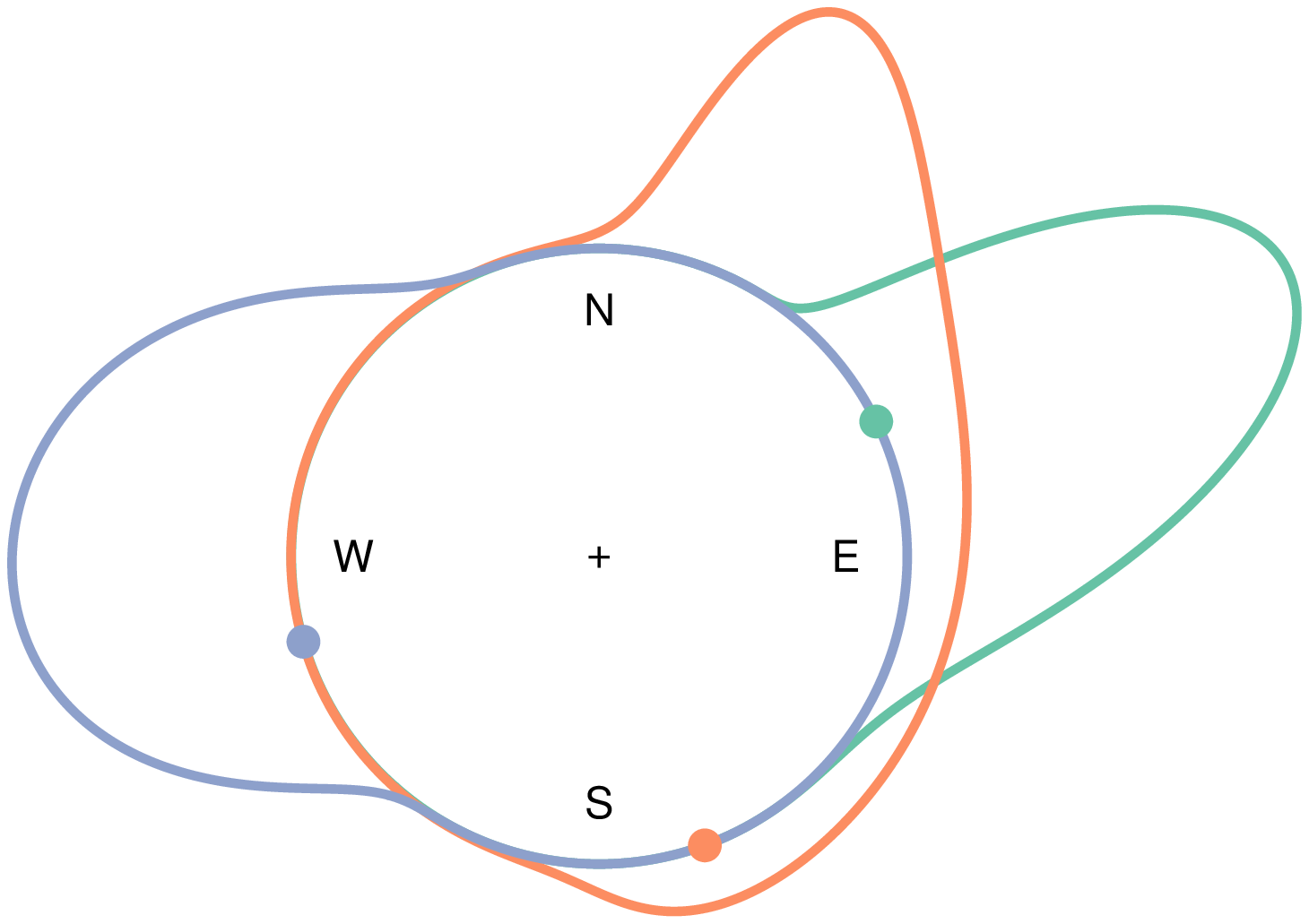}
    \caption{Illustration of the projected Gaussian process. Left:
      Realization of a projected Gaussian process with three
      highlighted locations (coloured arrows). Right: projected
      Gaussian densities at the three highlighted locations. Each blob
      corresponds to the observed directions}
    \label{fig:PGdistrib}
\end{figure}

Extending the above univariate distribution to the spatial setting
amounts to switch the bivariate vector $X$ for a bivariate Gaussian
process $\{X(s) = \{X_1(s), X_2(s)\}^\top\colon s \in \mathcal{X}\}$
with mean function $\mu\colon s \mapsto \mathbb{E}[\mathbb{X}(s)]$ and
covariance function $\gamma\colon (s_1, s_2) \mapsto \Gamma$ where
$\Gamma$ is a $4 \times 4$ block-matrix with each block
$\Gamma_{i,j} = \mbox{Cov}\{X(s_i), X(s_j)\}$, $i,j = 1, 2$. The
projected Gaussian process \citep{wangSpat} is then defined as
$\{\theta(s) = \arctan^*\{X_2(s) / X_1(s)\}\colon s \in
\mathcal{X}\}$, and will be denoted as
$\theta \sim PG\{\mu, \gamma\}$. Figure~\ref{fig:PGdistrib} plots one
realization of the projected Gaussian process and exhibits how it can
handle a large variety of behaviours, e.g., multi-modality, asymmetry.

Although no closed form exists for the projected Gaussian process
likelihood, inference becomes possible by working within an augmented
data framework. More precisely, for some spatial locations
$\mathbf{s} = (s_1, \ldots, s_k) \in \mathcal{X}^k$, $k \geq 1$, we
associate to the angular random vector
$\theta(\mathbf{s}) = \{\theta(s_1), \ldots, \theta(s_k)\}$ a radial
random vector $R(\mathbf{s}) = \{R(s_1), \ldots, R(s_k)\}$ so that one
can easily switch from polar coordinates
$\{R(\mathbf{s}), \theta(\mathbf{s})\}$ to Cartesian ones
$X(\mathbf{s}) = R(\mathbf{s}) \{ \cos \theta(\mathbf{s}), \sin
\theta(\mathbf{s})\}^\top$ and where multiplication is done
componentwise.

Inference is therefore based on the induced bivariate Gaussian random
vector $X(\mathbf{s})$ and it is easily seen that, for any
$\mathbf{r} \in (0, \infty)^k$ and $\mathbf{t} \in [0, 2\pi]^k$,
$k \ geq 1$, the joint density of the random vector
$\{R(\mathbf{s}), \theta(\mathbf{s})\}$ is
\begin{equation}
\label{eq:completedProjGaussDens}
f_{\mathbf{s}}(\mathbf{r}, \mathbf{t}) = (2\pi)^{-k/2}
|\Sigma(\mathbf{s})|^{-1/2} \exp\left( -\frac{\{\mathbf{r}^\top
    \mathbf{u} - \mu(\mathbf{s})\}^\top \Sigma(\mathbf{s})^{-1}
    (\mathbf{r}^\top\mathbf{u} - \mu(\mathbf{s})\}}{2}\right)
\prod_{i=1}^k r_i, 
\end{equation}
with
\begin{equation*}
    \mathbf{u} =
    \begin{bmatrix}
    \cos t_1 & \ldots & \cos t_k\\
    \sin t_1 & \ldots & \sin t_k
    \end{bmatrix}^\top,
\end{equation*}
where $\mu(\mathbf{s})$ and $\Sigma(\mathbf{s})$ are respectively the
mean vector and cross-covariance matrix of the Gaussian random vector
$X(\mathbf{s})$.

\section{Extreme--angular Bayesian hierarchical model}
\label{sec:extr-angul-bayes}

\begin{figure}
    \centering
    \begin{tikzpicture}[node distance={15mm},
    thick,
    roundnode/.style={circle, draw, very thick, minimum size=7mm},
    squarenode/.style={rectangle, draw, very thick, minimum size=5mm},
    scale=1]


\node[roundnode] (5)  {$m_\mu$};
\node[squarenode] (-5) [above of=5] {$\pi_{m_\mu}$};
\draw[->] (-5) -- (5);
\node[roundnode] (2) [below right of=5] {$\mu(s)$}; 
\node[roundnode] (6) [above right of=2] {$\gamma_{\mu}$};
\node[squarenode] (-6) [above of=6] {$\pi_{\gamma_\mu}$};
\draw[->] (-6) -- (6);
\draw[->] (5) -- (2); 
\draw[->] (6) -- (2);


\node[roundnode] (7) [right of=6] {$m_\sigma$};
\node[squarenode] (-7) [above of=7] {$\pi_{m_\sigma}$};
\draw[->] (-7) -- (7);
\node[roundnode] (3) [below right of=7] {$\sigma(s)$}; 
\node[roundnode] (8) [above right of=3] {$\gamma_{\sigma}$};
\node[squarenode] (-8) [above of=8] {$\pi_{\gamma_\sigma}$};
\draw[->] (-8) -- (8);
\draw[->] (7) -- (3); 
\draw[->] (8) -- (3);

\node[roundnode] (9) [right of=8] {$m_\xi$};
\node[squarenode] (-9) [above of=9] {$\pi_{m_\xi}$};
\draw[->] (-9) -- (9);
\node[roundnode] (4) [below right of=9] {$\xi(s)$};
\node[roundnode] (10) [above right of=4] {$\gamma_{\xi}$};
\node[squarenode] (-10) [above of=10] {$\pi_{\gamma_\xi}$};
\draw[->] (-10) -- (10);
\draw[->] (9) -- (4); 
\draw[->] (10) -- (4); 

\node (invisible1) at ($(2)!0.5!(3)$) {};
\node (invisible2) at ($(3)!0.5!(4)$) {};

\node[roundnode] (1) [below of = invisible1] {$\eta(s)$};
\draw[->] (2) -- (1);
\draw[->] (3) -- (1);
\draw[->] (4) -- (1);

\node[roundnode] (12) [below of = invisible2] {$X(s)$};
\node[roundnode] (14) [right= of 12] {$\{R(s),\theta(s)\}$};
\node[roundnode] (11) [below of=12] {$m_\theta$};
\node[squarenode] (-11) [below of=11] {$\pi_{m_\theta}$};
\draw[->] (-11) -- (11);
\node[roundnode] (13) [right of=11] {$\gamma_\theta$};
\node[squarenode] (-13) [below of=13] {$\pi_{\gamma_\theta}$};
\draw[->] (-13) -- (13);

\draw[->,double] (12) -- (14); 
\draw[->] (11) -- (12);
\draw[->] (13) -- (12);
\path[->]  (2) edge [bend right=60] node {}(12);
\draw[->] (3) -- (12);
\draw[->] (4) -- (12);
\end{tikzpicture}

\caption{Directed acyclic graph for the extreme--angular hierarchical
  model with the latent radial process
  $\{R(s)\colon s \in \mathcal{X}\}$. A double arrow indicates a
  deterministic relationship between the two nodes.}
    \label{fig:hierstru}
\end{figure}

By combining the latent variable for extremes~\eqref{eq:modelCooley}
and the projected Gaussian process described in
Section~\ref{sec:proj-gauss-proc}, one can define what we shall call a
spatial extremal--angular Bayesian hierarchical model:
\begin{align}
\label{eq:OurModel}
    \theta(\cdot)\mid \{m_\theta, \gamma_\theta, \mu(\cdot),
  \sigma(\cdot), \xi(\cdot)\} &\sim \mbox{PG}(m_\theta,
                                \gamma_\theta),\\ 
    \notag
    \eta(s) \mid \{\mu(s), \sigma(s), \xi(s)\}
                              &\stackrel{\text{ind}}{\sim}
                                \mbox{GEV}\{\mu(s), \sigma(s),
                                \xi(s)\}, \qquad s \in \mathcal{X}  \\ 
    \notag
      \mu(\cdot) \mid \{m_\mu, \gamma_\mu\} & \sim \mbox{GP}(m_\mu,
                                              \gamma_\mu)\\ 
      \notag
      \sigma(\cdot) \mid \{m_\sigma, \gamma_\sigma\} & \sim
                                                       \mbox{GP}(m_\sigma,
                                                       \gamma_\sigma)\\ 
      \notag
  \xi(\cdot) \mid \{m_\xi, \gamma_\xi\}& \sim \mbox{GP}(m_\xi, \gamma_\xi)
\end{align}
where prior distributions are put on the mean and covariance function
parameters on each (eventually projected) Gaussian processes. The last
four lines are identical to Model~\eqref{eq:modelCooley}. Further, and
as suggested by the conditioning terms, the mean function for the
angular process may depend on any relevant spatial covariates such as
longitude, latitude or elevation and, more importantly, on GEV
parameters. For example, a sensible choice may be to use, in addition
to geophysical covariates, GEV quantiles to assess whether the angular
may be impacted by the extreme values
intensities. Figure~\ref{fig:hierstru} gives the directed acyclic
graph for this model.

Due to the presence of latent processes $m_\mu(\cdot)$,
$m_\sigma(\cdot)$, $m_\xi(\cdot)$ and $R(\cdot)$, the likelihood has
an integral representation and, working within a Bayesian setting,
marginalization across the latent processes is done numerically using
MCMC techniques. The Gibbs sampler for Model~\eqref{eq:OurModel} is
similar to that proposed by \citet{Cooley} and \citet{wangSpat}, but,
due to the induced dependence between extremes and angles, some of the
full conditional distributions required for a Gibbs sampler are more
sophisticated. More precisely, updating the mean function parameters
of the Gaussian processes associated to the GEV distributions for
locations $\mathbf{s}$ consists in sampling from
\begin{equation}
\label{eq:fullConditionalMeanFctGEVLoc}
    \pi(m_\mu \mid \ldots) \propto \pi\{\mu(\mathbf{s}) \mid
    m_\mu(\mathbf{s}), \gamma_\mu(\mathbf{s}, \mathbf{s})\}
    \pi(m_\mu). 
\end{equation}
Note that, Equation~\ref{eq:fullConditionalMeanFctGEVLoc} corresponds
to that of Model~\ref{eq:modelCooley}. Similar expressions are found
for the mean function of the GEV scale and shape parameters as well as
for the parameters of the covariance functions $\gamma_\mu$,
$\gamma_\sigma$ and $\gamma_\xi$.

Contrary to Equation~\ref{eq:fullConditionalMeanFctGEVLoc} the update
of the GEV parameters at locations $\mathbf{s}$ differs and now
depends on the (augmented) projected Gaussian process. It consists in
sampling from
\begin{equation*}
    \pi\{\mu(s_j) \mid \ldots) \propto \prod_{i=1}^n \pi\{\eta_i(s_j)
    \mid \mu(s_j), \sigma(s_j), \xi(s_j)\} \pi\{X_i(\mathbf{s})\mid
    m_\theta, \gamma_\theta, \mu(\mathbf{s}), \sigma(\mathbf{s}),
    \xi(\mathbf{s}) \} \pi\{\mu(\mathbf{s}) \mid m_\mu, \gamma_\mu\}, 
\end{equation*}
for $j = 1, \ldots, k$ with similar expressions for $\sigma(s_j)$ and
$\xi(s_j)$, $j = 1, \ldots, k$.

Similarly, and since the radius process $R(\cdot)$ may depend on the
GEV parameters, the associated full conditional distributions differ
from that given in~\citet{wangSpat} and are given by
\begin{align*}
    \pi\{R_i(\mathbf{s}) \mid \ldots\} &\propto \pi\{R_i(\mathbf{s}) 
    \mid \theta_i(\mathbf{s}), m_\theta, \gamma_\theta,
                                         \mu(\mathbf{s}),
                                         \sigma(\mathbf{s}),
                                         \xi(\mathbf{s})\}, \qquad i =
                                         1, \ldots, n,\\ 
    \pi(m_\theta \mid \ldots) &\propto \prod_{i=1}^n
                                \pi\{X_i(\mathbf{s}) \mid m_\theta,
                                \gamma_\theta, \mu(\mathbf{s}),
                                \sigma(\mathbf{s}), \xi(\mathbf{s})\}
                                \pi(m_\theta), 
\end{align*}
where $\pi(m_\theta \mid \ldots)$ denotes the full conditional
distribution of the mean function and $\pi(m_\theta)$ is the
associated prior distribution. A similar expression for the cross
covariance function $\gamma_\theta$ can be found.

In practice, one usually assume some parametric expressions for the
mean and (cross) covariance functions of the latent Gaussian processes
and use conjugate prior distributions whenever possible. For instance
a sensible choice for the mean function is to assume a Gaussian prior
distribution and a linear form for the mean function, i.e.,
$m_j(\mathbf{s}) = D_j(\mathbf{s}) \boldsymbol{\beta}_j$,
$j \in \{\mu, \sigma, \xi, \theta\}$, where $D_j(\mathbf{s})$ is the
design matrix at location $\mathbf{s}$.

For the GEV latent processes, a parametric stationary and isotropic
covariance function can be used, e.g., a powered exponential
covariance
$\gamma(s, s + h) = \gamma(\|h\|) = \tau \exp\left\{-\left(\|h\| /
    \lambda\right)^\kappa\right\}$. Similarly to get efficient updates
for each sill parameters of the Gaussian processes, an inverse Gamma
prior distribution may be a relevant choice. Since no conjugate prior
distributions exist for the range and the shape parameters $\lambda$
and $\kappa$, a Metropolis--Hastings updating scheme has to be used.

Finally, and to ease inference, it is common practice to assume a
stationary isotropic and separable cross covariance function for the
bivariate Gaussian process $\{X(s)\colon s \in \mathcal{X}\}$, i.e.,
\begin{equation*}
    \gamma_\theta(s, s + h) = \gamma_\theta(\|h\|) = T \otimes \rho(\|h\|),  \qquad 
    T = 
    \begin{bmatrix}
    \tau_\theta & \rho_\theta \sqrt{\tau_\theta}\\
    \rho_\theta \sqrt{\tau_\theta} & 1
    \end{bmatrix}, \qquad \tau_\theta > 0, \quad \rho_\theta \in (-1, 1),
\end{equation*}
where $\otimes$ denotes the Kronecker product, $\rho$ is any
parametric correlation function, e.g., powered exponential,
$\tau_\theta = \mbox{Var}\{R(s) \cos \theta(s)\}$ and $\rho_\theta$ is
the cross--correlation, i.e.,
$\rho_\theta(s) = \mbox{Cor}\{R(s) \cos \theta(s), R(s) \sin
\theta(s)\}$, $s \in \mathcal{X}$. As suggested by \citet{wang}, we
set $\mbox{Var}\{R(s) \sin \theta(s)\} = 1$ to ensure identifiability
of the parameters. Explicit expressions for these full conditional
distributions are postponed to the Appendix.

\begin{algorithm}
    \SetKwInOut{Input}{input}\SetKwInOut{Output}{output}
    
    \Input{A Markov chain $\{\psi_t\colon t = 1, \ldots, N\}$ sampled
      from the Gibbs sampler introduced above and a new location $s_*
      \in \mathcal{X}$.} 
    
    \Output{Predictive posterior predictions of an unknown quantity
      $U(s_*)$.} 
    
    \BlankLine

    \For{$t=1, \ldots, N$}{
    \tcc{Conditional sampling of the GEV parameters}
    Sample $\mu_t(s_*)$ from the conditional distribution $\mu(s_*)
    \mid \mu(\mathbf{s}), \psi_t$\; 

    Sample $\sigma_t(s_*)$ and $\xi_t(s_*)$ in the same way\;

    \tcc{Conditional sampling for the angular component}
    Sample $\theta_t(s_*)$ from the conditional distribution
    $\theta(s_*) \mid \mu(\mathbf{s}), \sigma(\mathbf{s}),
    \xi(\mathbf{s}), \psi_t$\; 
    }

    Return the $U(s^*)$-estimator $\hat{U}(\psi_1(s_*), \ldots, \psi_N(s_*))$\;
    \caption{Pointwise predictive posterior predictions from the
      extreme--angular model.}
    \label{alg:prediction}
\end{algorithm}

Spatial models often aim at doing predictions at unobserved locations
$s_* \in \mathcal{X}$ such as the values of the GEV parameters, return
levels or the circular mode of the posterior distribution of
$\theta(s^*)$. Within a Bayesian framework, it is typically done
through the posterior predictive distribution
\begin{equation*}
\int \pi(\eta(s_*), \theta(s_*) \mid \psi, \mathcal{D}_n) \pi(\psi
\mid \mathcal{D}_n) \mbox{d$\psi$},
\end{equation*}
where
$\mathcal{D}_n = \{(\eta_i(s_j), \theta_i(s_j))\colon i = 1, \ldots,
n, j = 1, \ldots, k\}$ and $\boldsymbol{\psi}$ is the parameter vector
of the model. As expected, the above integral representation has no
closed form and, similarly to the Gibbs sampler introduced above, one
has to resort to numerical integration where each state of the
generated Markov chain is browsed, see
Algorithm~\ref{alg:prediction}. Further, when the unknown quantity
$U(s_*)$ has no closed form, one might use an empirical version of
$U(s_*)$ such as empirical quantiles of order $p$ to get a sensible
estimator.

\section{Simulation study}
\label{sec:simulation-study}

\begin{table}
  \caption{Configuration settings for the simulation study. The
    angular mean function is set to
    $m_\theta(s) = \{m_{\theta,1}(s), m_{\theta,2}(s)\}^\top$. For
    each configuration the latent GEV parameter processes
    $(\mu,\sigma,\xi)$ were held fixed respectively to Gaussian
    processes with mean functions
    $m_\mu(s)=2- 3 \text{lon}(s) - 2 \text{lat}(s)$,
    $m_\sigma(s)= 2 + \text{lat}(s)$, $m_\xi=0.05$, sill parameters
    $0.1,0.5,0.05$ and range parameters $1,2,3$ respectively.}
\label{tab:paramset}
\centering
\begin{tabular}{lccccc}
\hline
& $\sigma^2$ & $\rho$ & $m_{\theta,1}(s)$ & $m_{\theta,2}(s)$& Modality\\
\cline{2-6}
\phantom{II}I: Independent & 0.4 & 0.3 & $0.5$ & 0 & Multimodal\\
\phantom{I}II: Light & 0.4 & 0.3 & $0.5$& $ 2 \xi(s)$  & Multimodal\\
III: Strong & 1.0 & 0\phantom{.3} & $10 + 0.5 \mu(s)$ & 0 & Unimodal\\
\hline
\end{tabular}
\end{table}

In order to assess the performance of our algorithm, we run a
simulation study with three different dependence settings as shown in
Table~\ref{tab:paramset} and varying sample sizes $n$ and number of
locations $k$.  In Configuration I, the angular mean function
$m_\theta$ is independent of the GEV parameters and, consequently,
angles and extremes are assumed to be independent. With this setting,
Algorithm~\ref{alg:prediction} is equivalent to the use of two
independent samplers: one for the GEV component and one for the
angular component. Configuration II provides a linear dependence
between angles and extremes through the shape parameter of the
GEV. However, since the Gaussian process for the shape parameter has
constant mean and low variance, the shape parameter is roughly
constant over $\mathcal{X}$ and, as so, dependence between angles and
extremes magnitudes is limited. On the contrary, in Configuration III
the angular mean function depends on the GEV location parameter which
varies significantly over $\mathcal{X}$ yielding to a strong
dependence between angles and extremes magnitudes.

Working within a fixed domain framework, i.e., the spatial domain
$\mathcal{X}$ is fixed, we numerically analyse infill and sample size
asymptotics, i.e., where $k$ tends to infinity $n$ being fixed and
conversely. For each dependence setting of Table~\ref{tab:paramset},
we consider the cases where $k = 10, 25, 50$ and $n = 20, 50, 100$,
leading to 27 configurations overall.

\begin{figure}
    \centering
    \includegraphics[width=\textwidth]{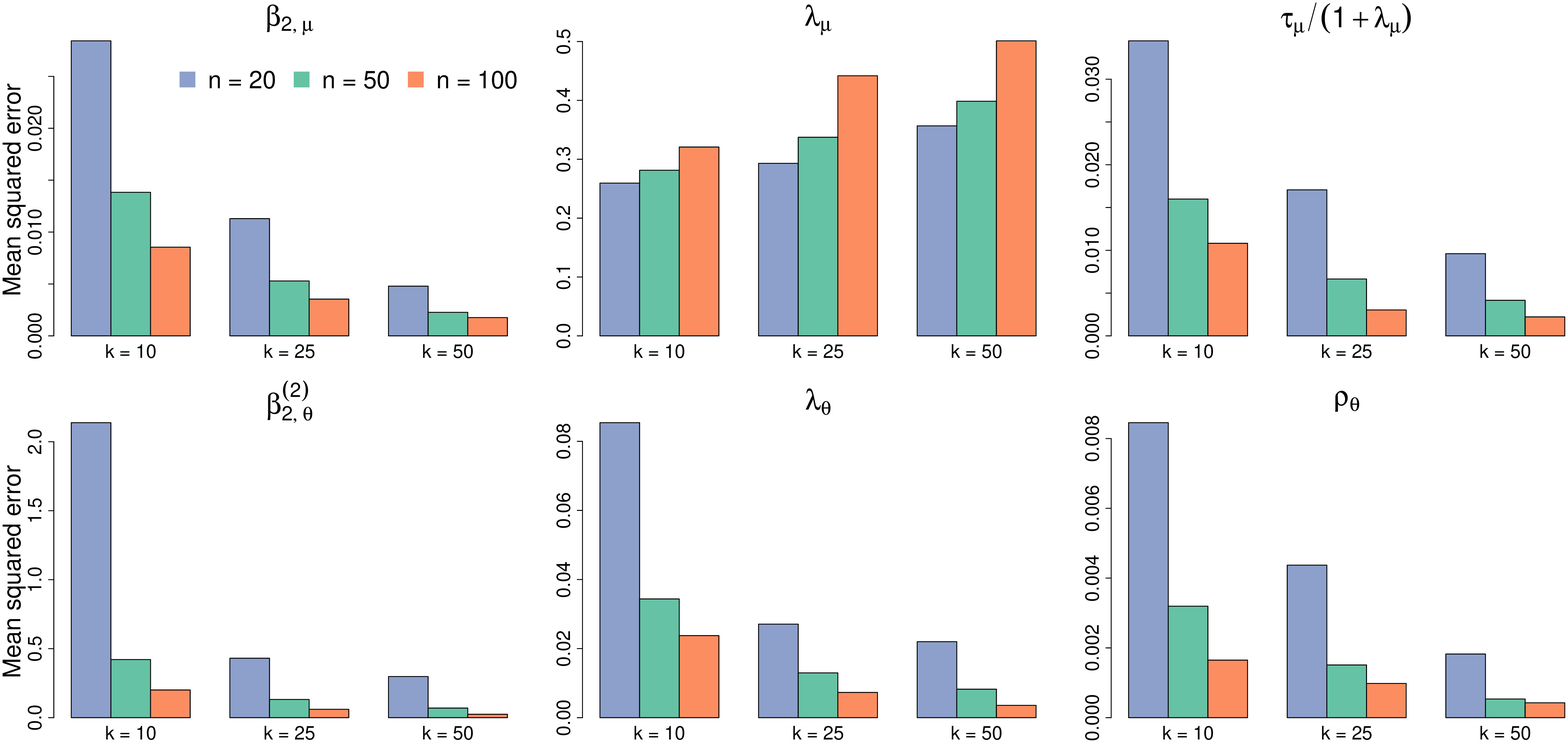}
    \caption{Evolution of the mean squared error (MSE) of parameters
      with varying number of observations $n$ and locations $k$ for
      Configuration II. Top: MSE for the regression (left), range
      (middle) parameters and quotient of scale and range (right) of
      the covariance function of the Gaussian process $\mu$. Bottom:
      MSE for regression parameter (left), range parameter (middle)
      and correlation parameter (right) of the projected Gaussian
      process $\theta$.}
    \label{fig:MSEparameters}
\end{figure}

Figure~\ref{fig:MSEparameters} shows the evolution of the mean squared
error as the number of locations $k$ and replicates $n$ vary for a
selected panel of parameters. The top row of
Figure~\ref{fig:MSEparameters} focuses on the Gaussian process of the
GEV location parameter. More precisely, results are shown for
$\beta_{2,\mu}$, a regression parameter of the mean function $m_\mu$,
the range parameter $\lambda_\mu$ of the covariance function
$\gamma_\mu$ as well as the ratio $\tau_\mu/(1+\lambda_\mu)$ where
$\tau_\mu$ is the sill parameter of $\gamma_\mu$. As expected, the
mean squared error for $\beta_{2,\mu}$ decreases as both $k$ and $n$
increase. Indeed, as $k$ increases, the number of GEV parameters
increases and the linear structure
$\mathbb{E}[\mu(s)] = x(s)^\top \boldsymbol{\beta}_\mu$ becomes more
apparent. Similarly, as the number of replicates $n$ increases, the
GEV parameters estimates become increasingly more accurate and the
above linear structure has less noise. Results for other parameters
and dependence configurations show similar patterns.

Interestingly, the mean squared error for $\lambda_\mu$ has a
completely different behaviour and no convergence is visible. A similar
pattern is seen for the sill parameter $\tau_\mu$. As shown
in~\citet{incEstimation}, there is no consistency in the estimation of
both the sill and range parameters of a Gaussian process with a single
realization. This result applies to our model as each location has a
single set of GEV parameters and, consequently, a single realization
of the Gaussian processes.

The bottom row of Figure~\ref{fig:MSEparameters} is similar to the top
row with an emphasis on the angular component, namely
$\beta_{2,\theta}^{(2)}$, $\rho_\theta$ and $\lambda_\theta$. As
expected and using the same arguments as the ones stated previously,
the evolution of the mean squared error $\beta_{2,\theta}^{(2)}$ is
similar to that for $\beta_{2,\mu}$. There is however a subtle
difference in the estimation of the parameters of the bivariate
Gaussian process $\{\boldsymbol{X}(s)\colon s \in \mathcal{X}\}$
compared to that for the GEV parameters, e.g.,
$\{\mu(s)\colon s \in \mathcal{X}\}$. Contrary to the latter case,
parameters are now estimated from $n > 1$ independent replicates and
no consistency issues exist.

\begin{figure}
    \centering
    \includegraphics[width=\textwidth]{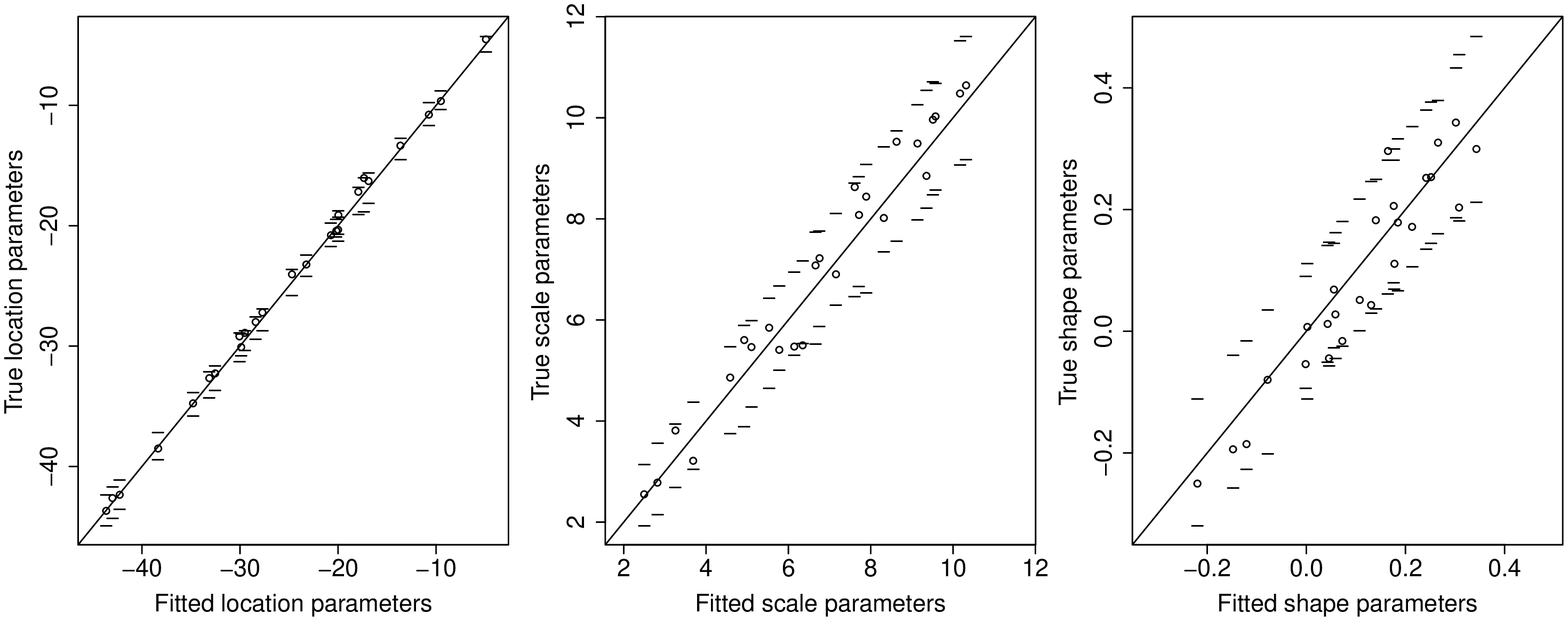}
    \caption{Posterior median (with 95\% credible intervals) for the
      GEV parameters at $k=25$ locations and with $n=50$ observations
      for Configuration II.}
    \label{fig:qqplot}
\end{figure}

At first sight, the above lack of consistency may cause problems in
the estimation of the GEV parameters or related quantities such as
quantiles. Fortunately, \citet{incEstimation} have shown that the
ratio of the sill and range parameters can be consistently estimated
(see top right panel of Figure~\ref{fig:MSEparameters}) and, more
importantly, that there is no impact on the predictions of the GEV
parameters. Figure~\ref{fig:qqplot} compares the true GEV parameters
from that estimated from our approach. As expected, one can see that
the predicted GEV parameters match the theoretical ones. The shape
parameter being usually harder to estimate, slightly worse
performances can be seen.

\begin{figure}
    \centering
    \includegraphics[width=\textwidth]{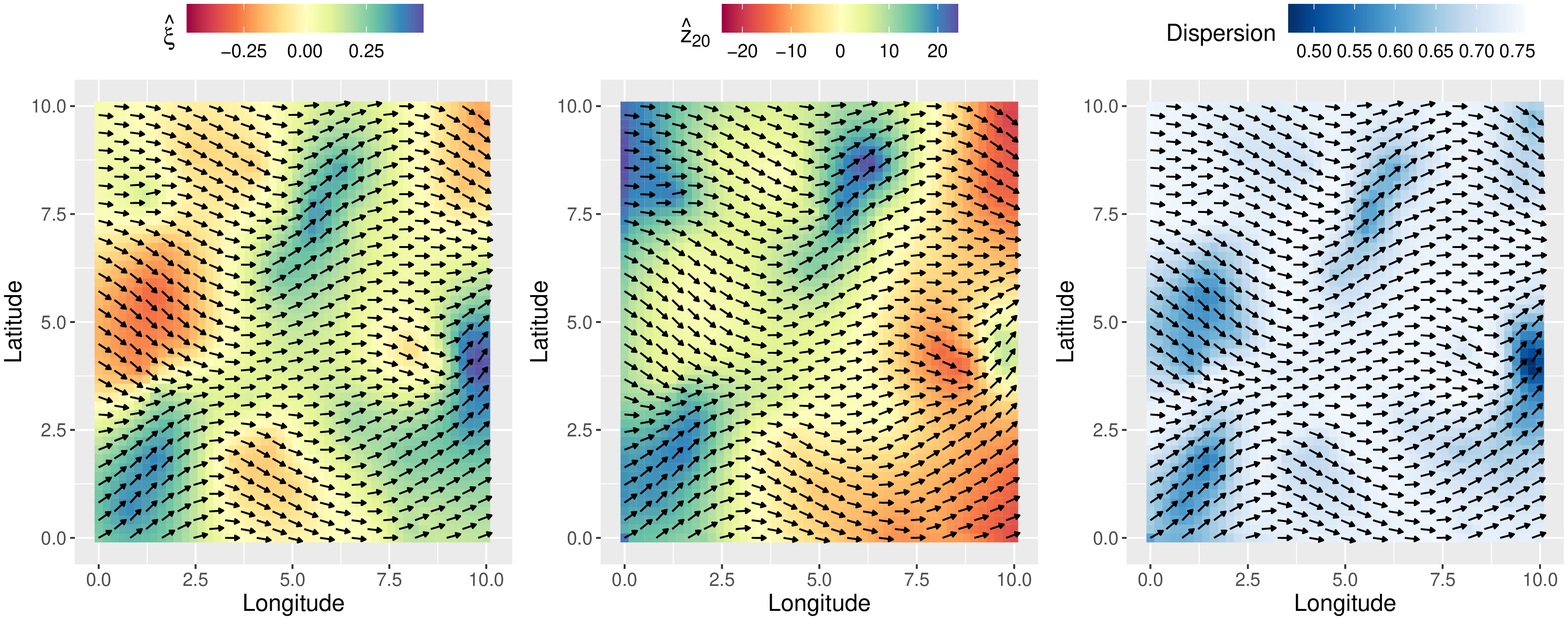}
    \caption{Prediction maps for Configuration II with $n=100$
      observations and $k=50$ locations. All predictions are derived
      from the predictive posterior distribution. Left: Prediction of
      the main direction and GEV shape parameter. Middle: Prediction
      of the main direction and GEV quantiles of order $0.95$. Right:
      prediction of the main direction and circular standard
      deviation.}
    \label{fig:mapsimul}
\end{figure}

We now assess the predictive performance of the proposed model and,
more specifically, that associated to the prediction of some angular
quantity over the spatial domain $\mathcal{X}$. As the projected
Gaussian process may be multimodal, some care is needed in defining
the quantity of interest and, in the sequel, we will focus on the
(main) posterior mode for the angle. Figure~\ref{fig:mapsimul} shows
prediction maps for Configuration II. With this setting, the
predictive posterior distribution is bimodal and may introduce spatial
discontinuities in the posterior modes. To be more specific, there is
a cutoff value for the shape parameter for which if exceeded, the
modal direction is top-right and bottom-right otherwise. The left
panel of Figure~\ref{fig:mapsimul} illustrates the relationship
between predicted angles and shape parameters as well as the
aforementioned cutoff behaviour. The middle panel is similar to the
previous one except that predicted quantiles of order 0.95 are now
overlaid. Since GEV quantiles are function of the GEV location, scale
and shape parameters, overall, the same behaviour can be seen. However,
due to the analytic expression for GEV quantiles, there is no cut--off
behaviour as the one stated previously. For instance, quantiles in the
outermost South--East sub-region are lower due to the contribution of
the location parameter as opposed to the general behaviour
``South--East direction for small return levels''. The right panel
shows the relationship between predicted angles and the angular
dispersion. One can see that the two modes have different impact. More
precisely, one can see that the mode associated to the North--East
direction is the preponderant one and has large dispersion whereas
that associated to the South--East direction only dominates after a
rare cut-off exceedance and has less less dispersion.

\section{Application}
\label{sec:application}

\begin{figure}
    \centering
    \includegraphics[width=0.33\textwidth]{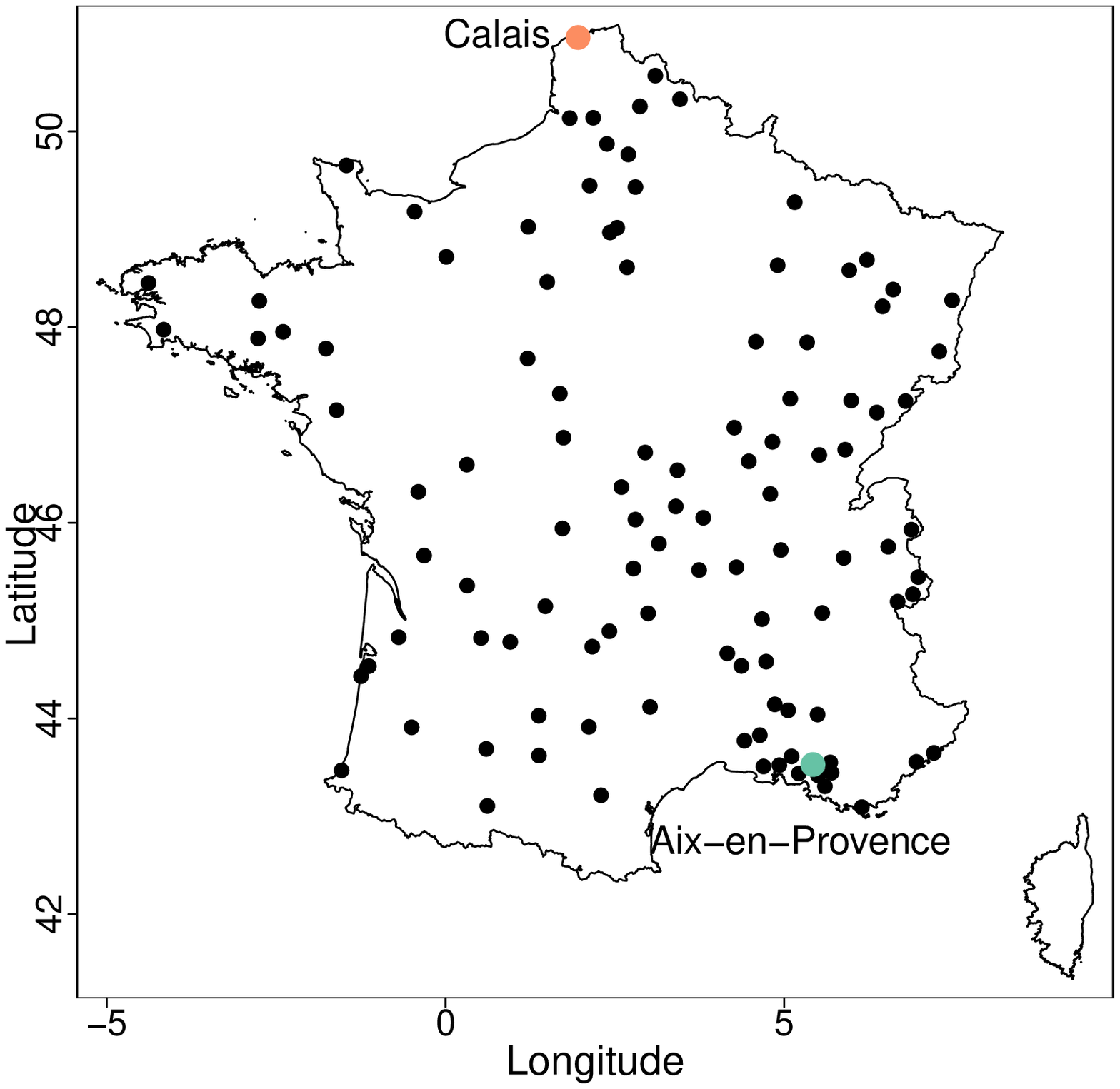}\hfill%
    \includegraphics[width=0.33\textwidth]{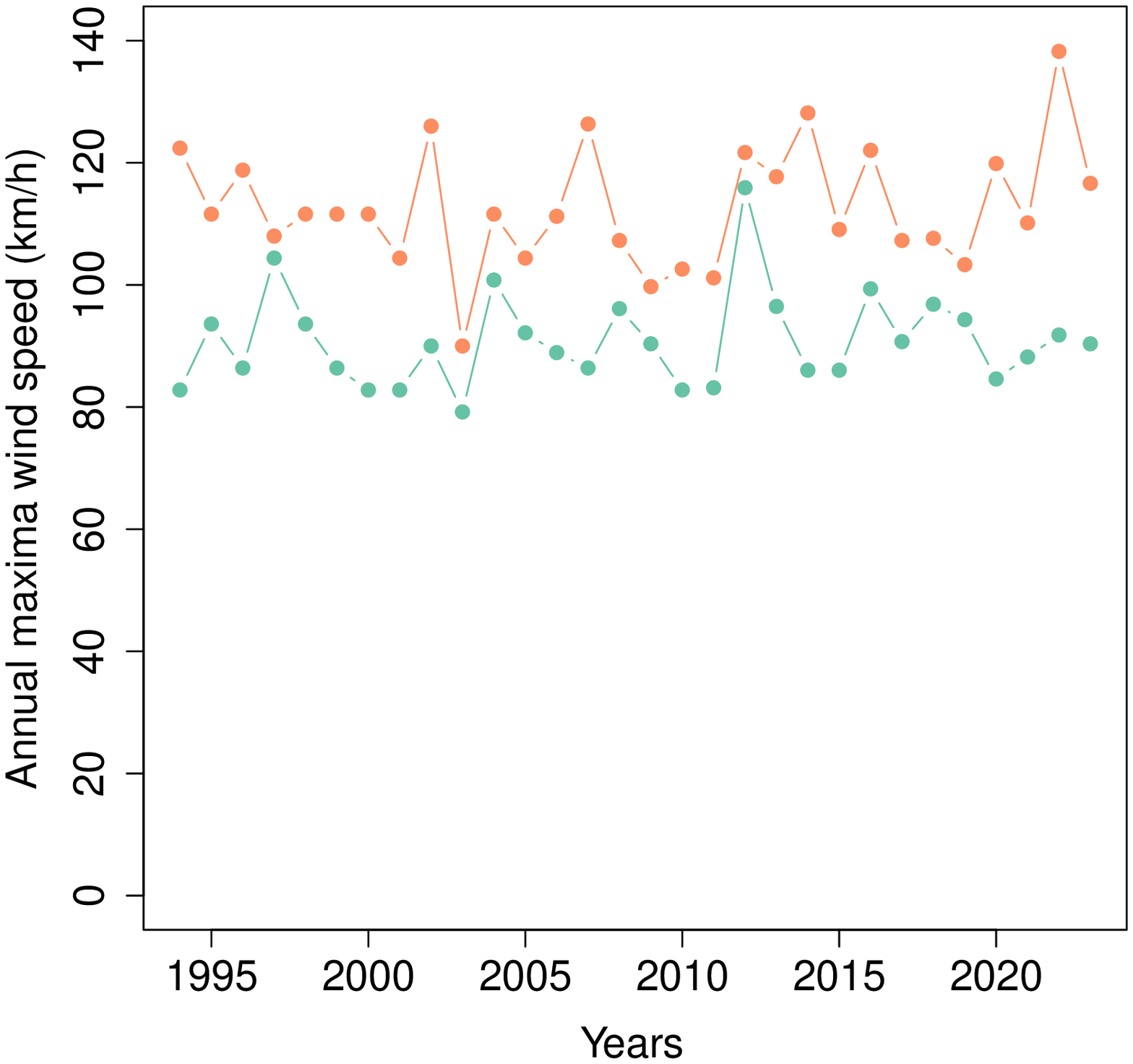}\hfill%
     \includegraphics[width=0.33\textwidth]{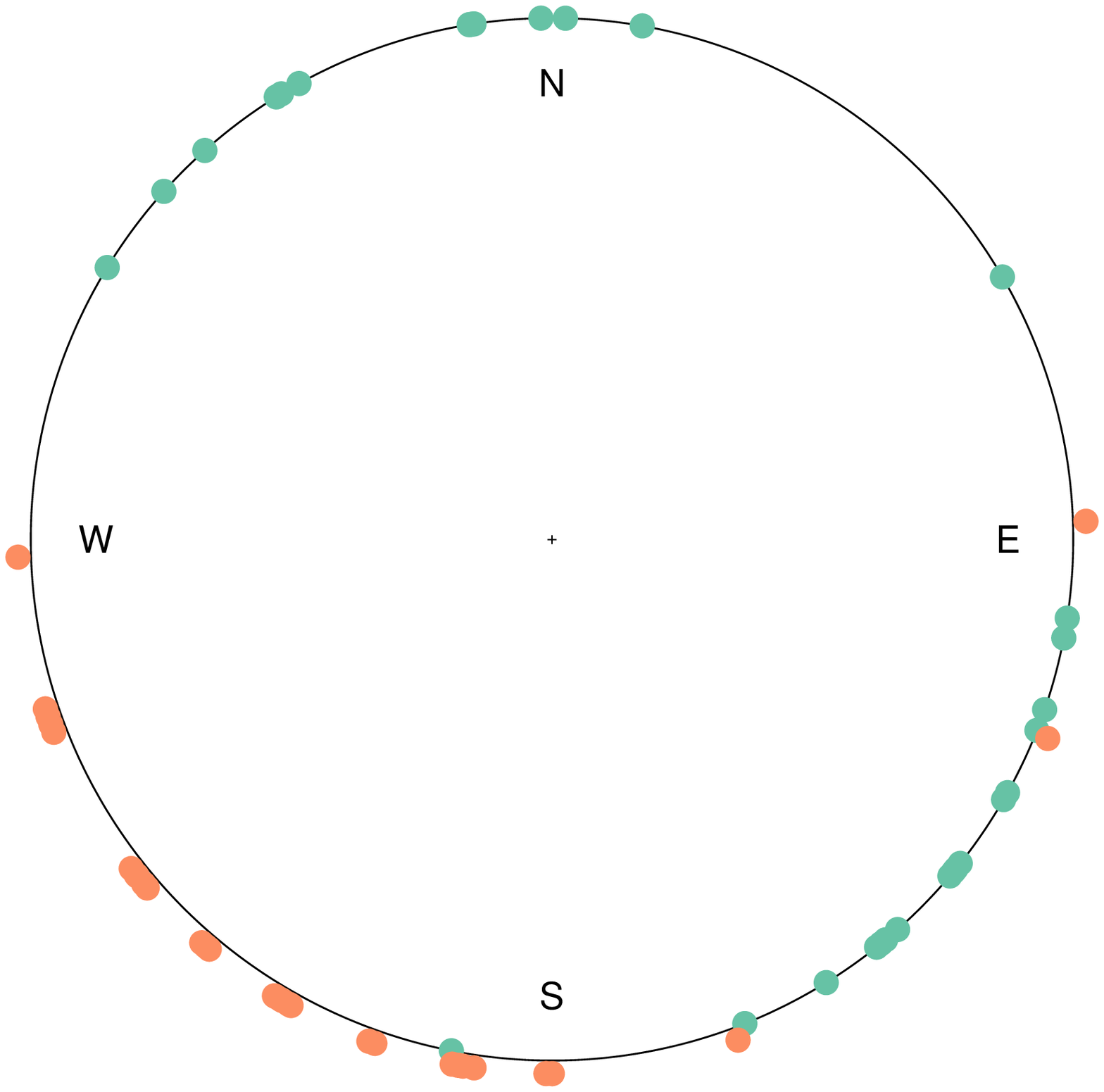}
     \caption{Wind data in continental France. From left to right:
       study region and locations of the weather stations; times
       series of annual maxima wind speed and empirical distribution
       of angles (direction of extreme wind) at two selected stations:
       Calais (orange) and
       Aix-en-Provence (green).}
    \label{fig:StudyRegion}
\end{figure}

The data, freely available from Météo-France, consist in annual maxima
of wind speeds and their associated wind direction observed at a
height of 10m for $k=110$ weather stations in France and recorded from
1994 to 2023. Figure~\ref{fig:StudyRegion} plots the spatial
distribution of the weather stations, the annual maxima wind speed
time series and the empirical distributions of wind directions
associated to the annual maxima for two selected stations. One can see
that annual wind speed maxima at Calais are larger than in
Aix-en-Provence and appear to have a single main direction
(South--West) while extreme wind speeds at Aix-en-Provence may
originate from two different directions (North--North--West and
South--East). As expected, the spatial distribution of extreme wind
speeds is not stationary over France; neither is that for wind
directions. Modelling such non-stationary behaviour is challenging,
but it is hoped that thanks to its flexibility, the use of
Model~\eqref{eq:OurModel} will be able to cope with those two
different types of non-stationarity.

\begin{table}
    \centering
    \caption{Widely Applicable Information Criterion (WAIC) for
      selected models. All models have $m_\xi(s) = \beta_0$. The
      function $q$ is the quantile function of the GEV distribution.}
    \label{tab:WAIC_models}
    \begin{tabular}{llccc}
    \hline
    & & WAIC$_\theta$ & WAIC$_\eta$ &WAIC  \\
    \hline 
      Model 0 &  & 14,702 & \textbf{16,343} & 31,045  \\
      & {$\begin{aligned}
        m_\mu(s) &= \beta_0 + \beta_1 \text{alt}(s)\\ 
     m_\sigma(s) &=\beta_0 + \beta_1 \text{lon}(s) + \beta_2 \text{lat}(s)  + \beta_3 \text{alt}(s)\\
    m_{\theta, 1}(s) &= \beta_0 + \beta_1 \text{lon}(s) + \beta_2 \text{lat}(s)  + \beta_3 \text{alt}(s)\\
     m_{\theta, 2}(s) &= \beta_0 + \beta_1 \text{lon}(s) + \beta_2
                        \text{lat}(s)  + \beta_3 \text{alt}(s)
      \end{aligned}$}
               &  &  &   \\ \\

      Model 1& Model 0 with   &  14,096 & 16,581 & 30,677  \\
    & {$\begin{aligned}
      m_{\theta, 1}(s) &= \beta_0 + \beta_1 \text{lon}(s) + \beta_2
                         \text{lat}(s)  + \beta_3 \text{alt}(s) +\\
      &\quad \beta_4 \mu(s) + \beta_5 \sigma(s)\\
      m_{\theta, 2}(s) &= \beta_0 + \beta_1 \text{lon}(s) + \beta_2
                         \text{lat}(s)  + \beta_3 \text{alt}(s) +\\
      &\quad \beta _4 \mu(s) + \beta_5 q(0.95 \mid \mu,
                         \sigma, \xi)
    \end{aligned}$}
       & & & \\ \\
       Model 2& Model 0 with   & 13,823 & 16,412 & 30,235 \\
     &        {$\begin{aligned}
         m_{\theta, 1}(s) &= \beta_0 + \beta_1 \text{lon}(s) + \beta_2
                            \text{lat}(s)+ \beta_3 \text{alt}(s) +\\
         &\quad \beta_4 q(0.95 \mid \mu, \sigma, \xi) + \beta_5 q(0.99 \mid \mu, \sigma, \xi)
       \end{aligned}$} & & & \\ \\
    \textbf{Model 3}& Model 2 with  & \textbf{13,800} & 16,352 & \textbf{30,152}    \\
    &   {$\begin{aligned}
      m_{\theta, 1}(s) &= \beta_0 + \beta_1  q(0.95 \mid \mu, \sigma,
                         \xi) + \beta_2 q(0.99 \mid \mu, \sigma, \xi)
    \end{aligned}
      $}  & & &\\ \\
    
    Model 4 & Model 2 with  & 13,822 & 16,364  & 30,186    \\
    &   {$\begin{aligned}
      m_\mu(s) &= \beta_0 \\
      m_\sigma(s) &= \beta_0+ \beta_1 \text{lat}(s)  +
                    \beta_2 \text{alt}(s)
    \end{aligned}
      $}  & & &\\ \\
     Model 5 & Model 3 with  & 13,858 & 16,418 & 30,276 \\
    &   {$\begin{aligned}
      m_\mu(s) &= \beta_0 \\
      m_\sigma(s) &= \beta_0+ \beta_1 \text{lat}(s)  + \beta_2
                    \text{alt}(s)
    \end{aligned}
      $} & & &\\
   
    \hline
    \end{tabular}
\end{table}

To perform model selection for Bayesian hierarchical models, a useful
performance metric is the Widely Applicable Information Criterion
(WAIC) \citep{WAIC} whose focus is on predictive performances. In the
same vein as the Akaike Information Criterion \cite{Akaike1974}, WAIC
penalizes model complexity but is more relevant for performing model
selection in a setting. Table~\ref{tab:WAIC_models} displays
performance scores for a bunch of competitive models with a varying
degree of complexity. Note that in addition to geophysical covariates
such as longitude, latitude and elevation, some models may also
consider extreme wind speed quantiles for the modelling of wind speed
direction. The rationale for this type of dependence is that it may be
sensible to assert that the largest extreme wind speeds are attached
to a very specific wind direction.

Although a large number of models have been considered, and to save
space, results are only presented for a limited number of
configurations. Model 0 assumes independence between the extreme value
process $\{\eta(s)\colon s \in \mathcal{X}\}$ and the angular process
$\{\theta(s)\colon s \in \mathcal{X}\}$ while all the other models
assume dependence. One can see that the independent model performs
best in predicting extreme wind speeds but is poor in predicting wind
direction. Hence, for this application, it seems that the knowledge of
wind direction is irrelevant in modelling wind speeds. Model 3 appears
to be the most accurate as it is almost as efficient as the
independent model in predicting wind speed and has good predictive
performances for wind directions. This model makes use of
the 20 and 100--years return levels and may indicate that the original
assumption that the largest extreme wind speeds impact wind directions
is sensible.

\begin{figure}
    \centering
    \includegraphics[width=0.33\textwidth]{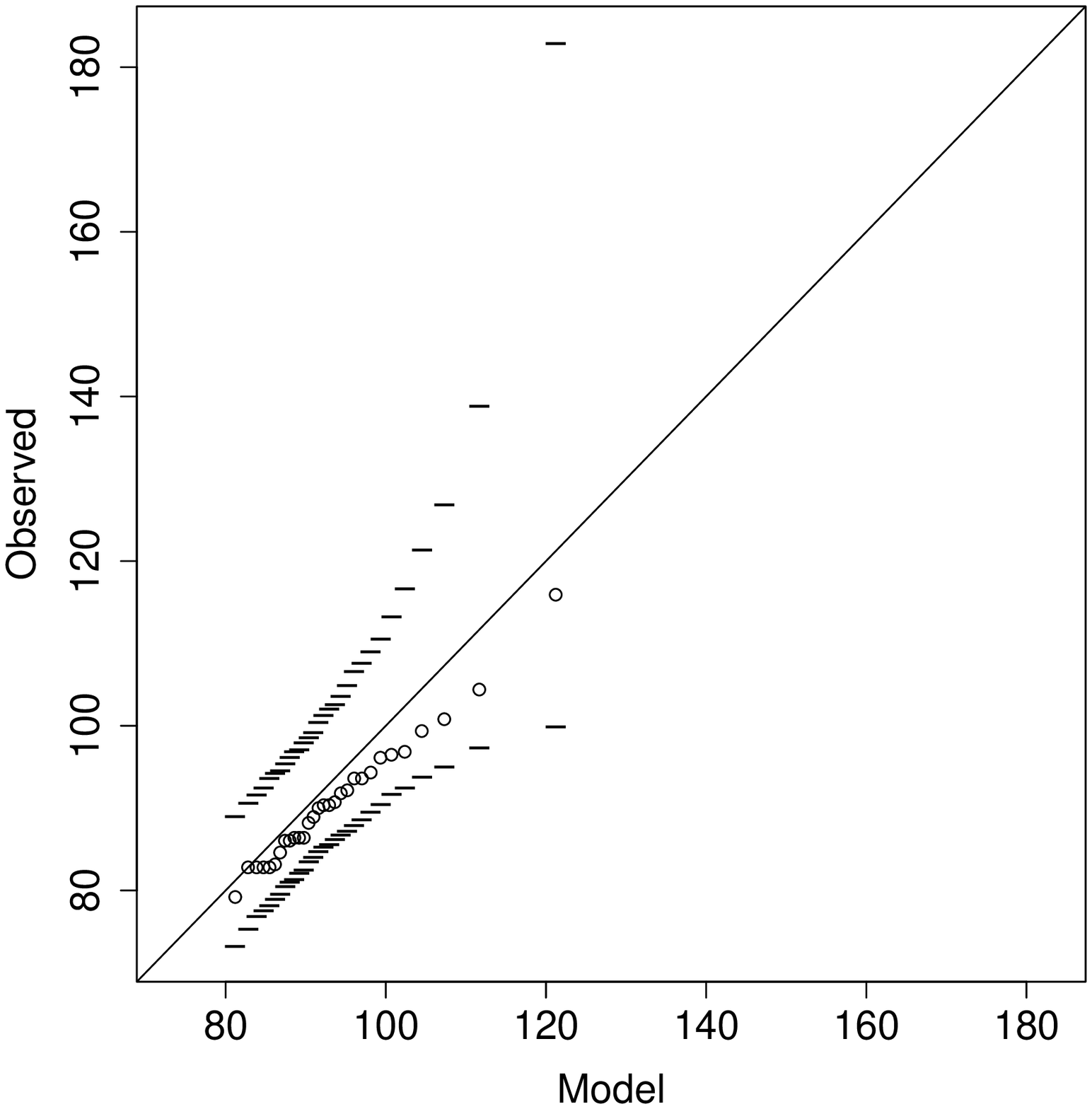}\hfill%
    \includegraphics[width=0.33\textwidth]{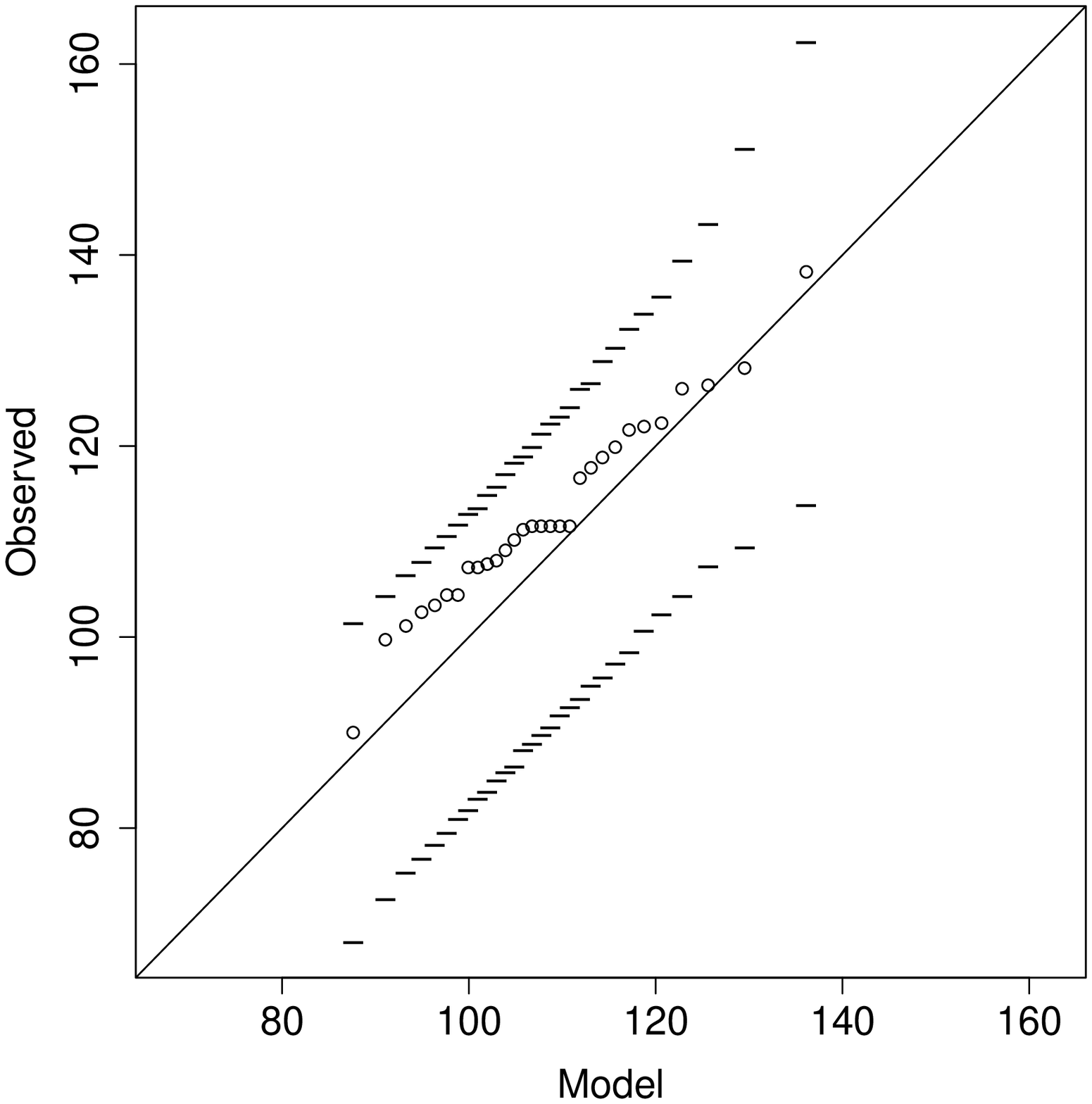}\hfill%
    \includegraphics[width=0.33\textwidth]{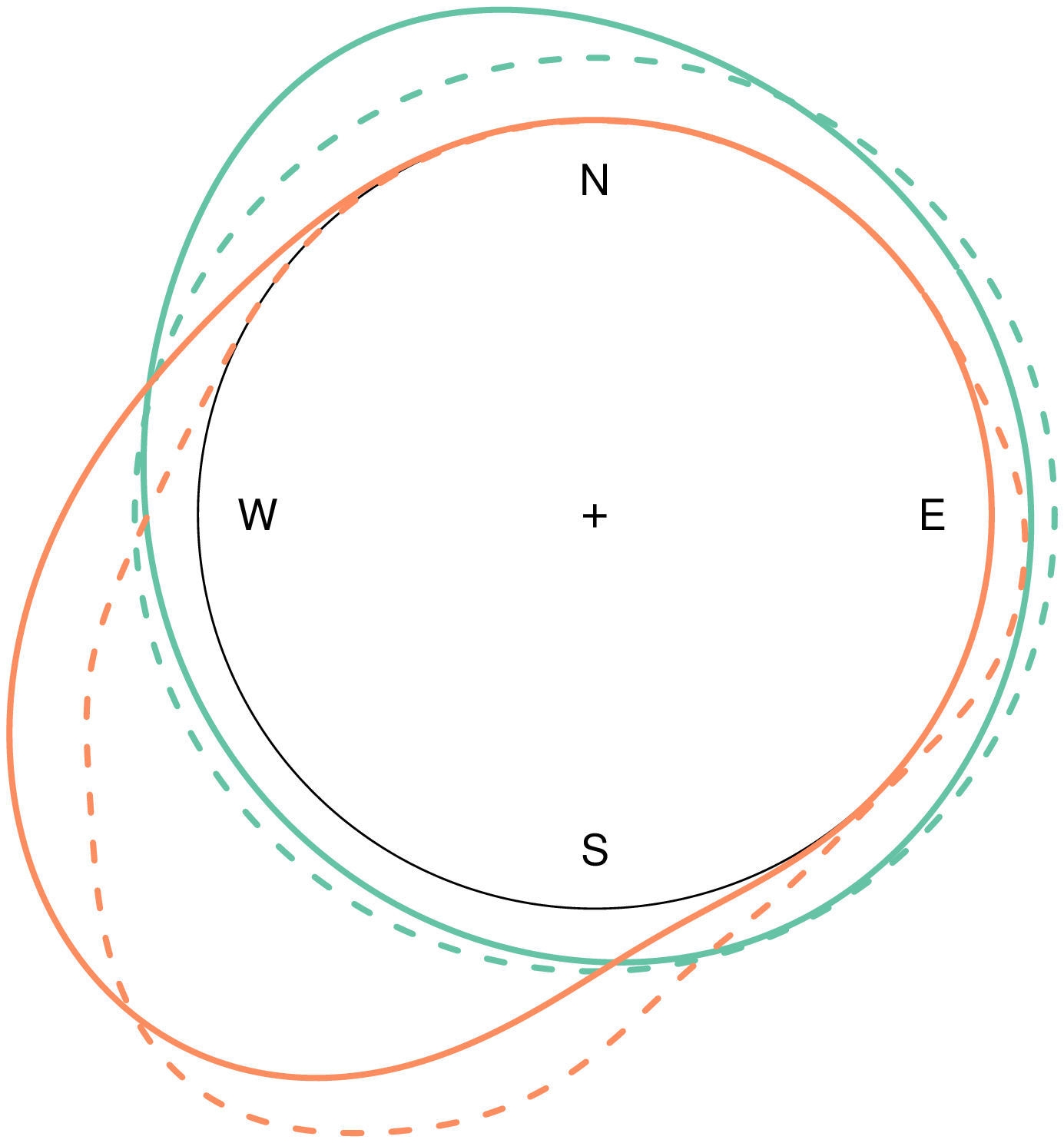}
    \caption{Model checking. The left and middle panels compare the
      observed and predicted maxima for the Aix-en-Provence and Calais
      stations (with 95\% confidence envelopes). The right panel
      compares the predicted angular distributions (solid lines) and
      the angular kernel density estimates (dashed lines).}
    \label{fig:verif_valid}
\end{figure}

Figure~\ref{fig:verif_valid} assesses the performance of Model~3. The
left panel compares predicted observations and that observed wind
speeds for the two highlighted stations of
Figures~\ref{fig:StudyRegion}.  Although predictions show some bias,
predictions are rather good. Better performances are likely to be
obtained by using additional geophysical covariates, but
unfortunately, such additional covariates were not provided in the
data set who was limited to longitude, latitude and altitude only. The
right panel compares the observed directions and the fitted angular
densities for these two locations. Predictions appear to be accurate
and one can see that our model is able to cope with both unimodal
(Calais) and bi-modal (Aix-en-Provence) distributions. Similar results
were obtained for all the other stations.

\begin{table}
    \centering
    \caption{Posterior median and 95\% credible intervals (in
      parenthesis) for Model 3.}
    \label{tab:estimates_application}
    {\scriptsize
    \begin{tabular}{lccccccc}
    \cline{1-8}
     \multicolumn{8}{c}{Generalized extreme value layer}\\
    & $\beta_0$& $\beta_\text{lon}$& $\beta_\text{lat}$ & $\beta_\text{alt}$ & $\tau$ & $\lambda$ &\\
      \cline{1-7}
    $\mu$ & $95_{\tiny(93,97)}$&--- & --- & $2.4_{\tiny(-2.4,           7.3)}$ &
                                                            $78_{\tiny(60,102)}$
                                                                                      &
                                                                                        $8_{\tiny(1,20)}$ &\\ 
    $\sigma$ &  $2_{\tiny(-13,18)}$ &                              $-0.3_{\tiny(-0.7,0.1)}$
                                   & $0.19_{\tiny(-0.15,0.53)}$ &
                                                                   $2.9_{\tiny(1.2,4.9)}$
                                                                             &
                                                                               $6.5_{\tiny(4,11)}$ & $104_{\tiny(57,184)}$ & \\ 
    $\xi$ & $-0.07_{\tiny(-0.1,0)}$& ---& ---& --- &
                                                          $0.01_{\tiny(0.00,0.01)}$
                                                                                      &
                                                                                        $127_{\tiny(75, 225)}$ &\\ 
       \cline{1-8}
      \cline{1-8}
    &&&&&&&\\[0.05em]
    \multicolumn{8}{c}{Angular layer}\\
    \hline
    \multirow{2}{*}{ $m_{\theta,1}$} & $\beta_0$ &  $\beta_\text{q(0.95)}$& $\beta_\text{q(0.99)}$ & & $\tau$ & $\lambda$  &  $\rho_\theta$ \\ 
    & $-1.7_{\tiny(-3,-0.4)}$ & $-0.20_{\tiny(-0.26,-0.15)}$ &
                                                                 $0.18_{\tiny(0.14,0.23)}$ & & \multirow{3}{*}{$0.6_{\tiny(0.5,0.7)}$}
                                                                             &
                                                                               \multirow{3}{*}{$45_{\tiny(41,49)}$} &\multirow{3}{*}{$-0.3_{\tiny(-0.4,-0.2)}$}  \\  
    \cline{1-5}    \multirow{2}{*}{ $m_{\theta,2}$}& $\beta_0$ &
                                                                 $\beta_\text{lon}$&
                                                                                     $\beta_\text{lat}$ & $\beta_\text{alt}$  & &  &   \\ 
      &  $8.4_{\tiny(7.1,9.7)}$ & $0.03_{\tiny(0.01,0.05)}$  &
                                                                 $-0.19_{\tiny(-0.22,-0.16)}$
                                                        &
                                                          $-0.2_{\tiny(-0.3,
                                                          0.0)}$ &  &  & \\ 
    \hline
    \end{tabular}
    }
\end{table}

Table~\ref{tab:estimates_application} shows parameter estimates for
Model~3. Interestingly, the sum between the parameter estimates
associated to the 20 and 100--years return levels varies around 0,
i.e., $\beta_{q(0.95)} + \beta_{q(0.99)} \approx 0$.  As
$q(p_1) - q(p_2) = \sigma c(p_1, p_2, \xi)$ where $q(p)$ is the
quantile function of a $\mbox{GEV}(\mu, \sigma, \xi)$ and
$c(p_1, p_2, \xi)$ only depends on $p_1, p_2$ and $\xi$, extreme wind
speed direction appear to depend on the spread, skewness and higher
order moments of extreme wind speeds but not their central tendency.

\begin{figure}
    \centering
     \includegraphics[width=\textwidth]{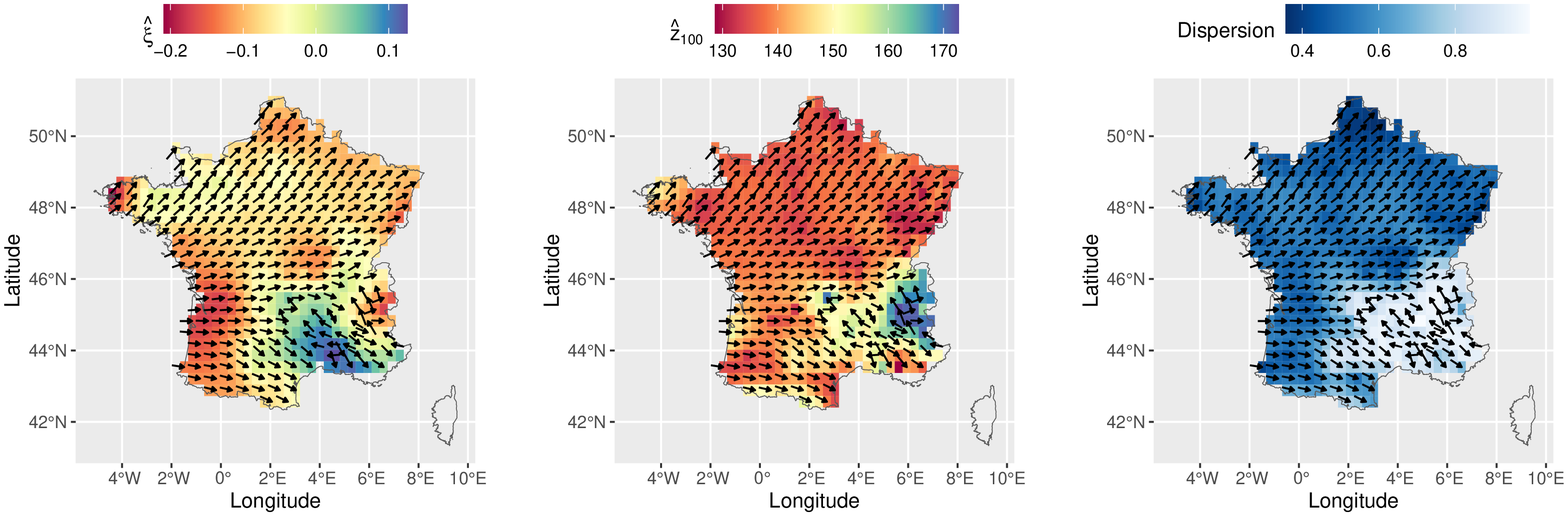}\\%
    \caption{Pointwise prediction maps of the wind direction and the
      GEV shape parameter (left), 100--year return level (middle) and
      the wind direction dispersion (right).}
    \label{fig:mapsimul_app}
\end{figure}

Figure~\ref{fig:mapsimul_app} gives predictions maps from Model~3. One
can see that the shape parameter appears to be rather constant over
France apart from the Atlantic coastline and the South--East part of
France. Similarly, the 100--years return levels are larger in the most
Western part, more precisely in the western part of Brittany and
Landes as well as the South--East part of France---especially in the
French Alps. Overall, the wind direction dispersion is rather
limited. One exception is the South--East part of France where the
angular dispersion is much larger indicating that wind speed
directions may vary roughly over space in that region. Further, as
expected, one can see that extreme wind event appears to come from
West but tends to change smoothly to the South West direction for the
Northern part of France. Again the South--East part of France and the
Alps show a completely different behaviour. For the former, extreme
wind events appear to come from West--North and is in agreement with
the observational study of \cite{Mistral}. The behaviour in the French
Alps is more erratic which may be explained that the wind trajectory
has to follow the valley in mountainous regions.

\section*{Conclusion}
\label{sec:conclusion}

Since many environmental processes are associated with an angular
component, this paper proposed a Bayesian hierarchical model to handle
such features by allowing the joint modelling of extremes and angular
components. An inferential framework has been proposed whose
performances were assessed through a simulation study. A lack of
consistency for some Gaussian processes dependence parameters has been
illustrated but, fortunately, did not impact prediction of relevant
quantities of interest such as return levels. An application to
extreme wind speeds and direction in continental France was
conducted. Results show that extreme wind speeds in the Atlantic part
of France is mainly dominated by West directions. A different behaviour
is seen for the Mediterranean part of France where extreme winds arise
from multiple winds patterns, i.e., north and south
directions. This leads to a multimodal angular distribution, whose
low-scale spatial variations are harder to recover with a general
model.

Although this paper illustrates the usefulness of the proposed model,
it has some limitations. First, the model is time stationary and
cannot handle situations where there is a clear impact of climate
change. An extension to the non-stationary case may be considered
using a non--stationary covariance structure or time dependent model
parameters. Second, as a consequence of the conditional independence
assumption, areal quantities such as total amount of rainfall in a
sub-domain is not possible. To overcome this problem, the conditional
independence assumption should be relaxed and, in accordance to the
extreme value theory, a max-stable should now be considered. The use
max-stable process is however challenging since the likelihood
associated to these processes is extremely CPU demanding and one has
to develop an elegant framework to overcome this computational burden.

\appendix
\section*{Appendix: Gibbs sampler}
\label{sec:appendixA}

Inference for our latent extreme--angular Bayesian hierarchical model
may be performed using a Gibbs sampler, whose steps we now
describe. To ease notations, we define
$\boldsymbol{R}_t = \{R_t(s_1), \ldots, R_t(s_k)\}$ and
$\boldsymbol{\mu}_t = \{\mu_t(s_1), \ldots, \mu_t(s_k)\}$ with similar
for the GEV scale and shape parameters. Given a current value of the
Markov chain
\begin{equation*}
    \psi_t = (\boldsymbol{R}_t,\boldsymbol{\mu}_t,
    \boldsymbol{\sigma}_t, \boldsymbol{\xi}_t, \rho_t, \tau_{\mu},
    \tau_{\theta, t}, \tau_{\sigma, t}, \tau_{\xi, t},
    \lambda_{\theta, t}, \lambda_{\mu, t}, \lambda_{\sigma, t},
    \lambda_{\xi, t}, \boldsymbol{\beta}_{\theta, t},
    \boldsymbol{\beta}_{\mu, t}, \boldsymbol{\beta}_{\sigma, t},
    \boldsymbol{\beta}_{\xi, t}),  
\end{equation*}
the next state $\psi_{t+1}$ of the chain is obtained as
follows.\medskip

\noindent
\textbf{Step 1: Updating the GEV parameters at each site}\\
Each component of $\boldsymbol{\mu} = \{\mu(s_1), \ldots, \mu(s_k)\}$
is updated singly according to the following scheme.  Generate a
proposal $\mu_p(\mathbf{s})$ from a symmetric random walk and compute
the acceptance probability
\begin{equation*}
  \alpha\{\mu(\mathbf{s}), \mu_p(\mathbf{s})\} = \min \left\{1, 
  r_1\{\mu(\mathbf{s}), \mu_p(\mathbf{s})\} r_2\{\mu(\mathbf{s}),
  \mu_p(\mathbf{s})\} r_3\{\mu(\mathbf{s}), \mu_p(\mathbf{s})\}
\right\}, 
\end{equation*}
with
\begin{align*}
    r_1\{\mu(\mathbf{s}), \mu_p(\mathbf{s})\} &= \prod\limits_{j=1}^k
                                                \frac{\pi\{\eta(s_j)
                                                \mid \mu_p(s_j),
                                                \sigma(s_j),
                                                \xi(s_j)\}}{\pi\{\eta(s_j)
                                                \mid \mu(s_j),
                                                \sigma(s_j),
                                                \xi(s_j)\}}\\ 
    r_2\{\mu(\mathbf{s}), \mu_p(\mathbf{s})\} &= \prod\limits_{i=1}^n
                                                \frac{\pi\{R_{i}(\mathbf{s}),
                                                \theta_i(\mathbf{s})
                                                \mid
                                                \mu_p(\mathbf{s}),
                                                \sigma(\mathbf{s}),
                                                \xi(\mathbf{s}),
                                                \boldsymbol{\beta}_{\theta},
                                                \rho, \tau_{\theta,
                                                t}\}}{\pi\{R_ 
    {i}(\mathbf{s}), \theta_i(\mathbf{s}) \mid \mu(\mathbf{s}),
                                                \sigma(\mathbf{s}),
                                                \xi(\mathbf{s}),
                                                \boldsymbol{\beta}_{\theta},
                                                \rho, \tau_{\theta,
                                                t}\}}\\ 
    r_3\{\mu(\mathbf{s}), \mu_p(\mathbf{s})\} &=
                                                \frac{\pi(\boldsymbol{\mu}_p
                                                \mid
                                                \boldsymbol{\beta}_\mu,
                                                \tau_\mu,
                                                \lambda_\mu)}{\pi(\boldsymbol{\mu}
                                                \mid
                                                \boldsymbol{\beta}_\mu,
                                                \tau_\mu,
                                                \lambda_\mu)} 
\end{align*}
where $r_1$ is a ratio of GEV likelihoods, $r_2$ a ratio of projected
Gaussian likelihoods and $r_3$ a ratio of multivariate Gaussian
likelihoods. With probability
$\alpha\{\mu(\mathbf{s}), \mu_p(\mathbf{s})\}$, the $\mu(\mathbf{s})$
component of $\boldsymbol{\psi}_{t+1}$ is set to $\mu_p(\mathbf{s})$;
otherwise it remains at $\mu(\mathbf{s})$. The scale and shape
parameters are updated similarly.

Due to possible components related to $\mu, \sigma$ or $\xi$, the
design matrix $D_\theta$ (related to the regression parameter
$\beta_\theta$) needs to be updated each time one of those parameters
is changed.  \medskip

\noindent
\textbf{Step 2: Updating the radius at each site and replicate}\\
Components of $\mathbf{R}_{i} = \{R_{i}(s_1), \ldots, R_{i}(s_k)\}$,
$i=1, \ldots, n$, are updated one by one according to the following
scheme. Generate a proposal $R_{p,i}(s_j)$ from a log-normal
distribution and accept with probability
\begin{equation*}
  \alpha\{R_{i}(s_j), R_{p,i}(s_j)\} = \min \left\{1, 
    \frac{\pi(\mathbf{R}_{p,i}, \boldsymbol{\theta}_i \mid
      \boldsymbol{\mu}, \boldsymbol{\sigma}, \boldsymbol{\xi},
      \boldsymbol{\beta}_{\theta}, \tau_{\theta, t}, \rho,
      \lambda_{\theta})}{\pi(\mathbf{R}_{i}, \boldsymbol{\theta}_i
      \mid \boldsymbol{\mu}, \boldsymbol{\sigma}, \boldsymbol{\xi},
      \boldsymbol{\beta}_{\theta}, \tau_{\theta, t}, \rho,
      \lambda_{\theta})}\right\}, 
\end{equation*}
i.e., a ratio of radial Gaussian likelihoods, based
on~\eqref{eq:completedProjGaussDens}.

\noindent
\textbf{Step 3: Updating the angle regression parameters}\\
Due to the use of conjugate priors, $\boldsymbol{\beta}_{\theta}$ is
drawn directly from a multivariate Gaussian distribution having
covariance matrix and mean vector
\begin{equation*}
 \{(\Sigma_{\theta}^*)^{-1} + nD_\theta^T \Sigma_{\theta}^{-1}
 D_\theta\}^{-1}
,\quad   \{(\Sigma_{\theta}^*)^{-1} + D_\theta^T \Sigma_{\theta}^{-1}
 D_\theta\}^{-1} \left\{(\Sigma_\theta^*)^{-1} \mu_{\theta}^* +
  D_\theta^T \Sigma_{\theta}^{-1} \sum\limits_{i=1}^n\mathbf{X}_{i} \right\}
\end{equation*}
where $\mu_{\theta}^*$ and $\Sigma_{\theta}^*$ are the mean vector and
covariance matrix of the prior distribution for
$\boldsymbol{\beta}_\mu$, $D_\theta$ is the design matrix related to
the regression coefficients $\boldsymbol{\beta}_\theta$,
$\mathbf{X}_{i}$ is the vector
$\begin{pmatrix} \mathbf{R}_{i}\cos(\boldsymbol{\theta}_i) &
  \mathbf{R}_{i}\sin(\boldsymbol{\theta}_i)
\end{pmatrix}^\top $ and $\Sigma_{\theta}$ the covariance matrix of $\mathbf{X}_{i}$. \medskip

\noindent
\textbf{Step 4: Updating the GEV regression parameters}\\
Due to the use of conjugate priors, $\boldsymbol{\beta}_{\mu}$ is
drawn directly from a multivariate Gaussian distribution having
covariance matrix and mean vector
\begin{equation*}
 \{(\Sigma_{\mu}^*)^{-1} + D_\mu^T \Sigma_{\mu}^{-1}
  D_\mu\}^{-1}
,\quad   \{(\Sigma_{\mu}^*)^{-1} + D_\mu^T \Sigma_{\mu}^{-1}
  D_\mu\}^{-1} \{(\Sigma_\mu^*)^{-1} \mu_{\mu}^* +
  D_\mu^T \Sigma_{\mu}^{-1} \boldsymbol{\mu} \}
\end{equation*}
where $\mu_{\mu}^*$ and $\Sigma_{\mu}^*$ are the mean vector and
covariance matrix of the prior distribution for
$\boldsymbol{\beta}_\mu$, $D_\mu$ is the design matrix related to the
regression coefficients $\boldsymbol{\beta}_\mu$ and $\Sigma_{\mu}$
the covariance matrix of $\boldsymbol{\mu}$. Again the regression
parameters for the GEV scale and shape parameters are updated
similarly.\medskip

\noindent
\textbf{Step 5: Updating the sill parameters of the covariance
  function}\\
$\tau_{\mu}$ is drawn directly from an inverse Gamma distribution
whose shape and rate parameters are
\begin{equation*}
  \frac{k}{2} + \kappa_{\tau_\mu}^*,\quad 
  \theta_{\tau_\mu}^* + \frac{1}{2} \tau_{\mu} (\boldsymbol{\mu} -
  D_\mu \boldsymbol\beta_{\mu})^T \Sigma_{\mu}^{-1}
  (\boldsymbol{\mu} - D_\mu \beta_{\mu}),
\end{equation*}
where $\kappa_{\tau_\mu}^*$ and $\theta_{\tau_\mu}^*$ are respectively
the shape and scale parameters of the inverse Gamma prior
distribution. The sill parameters of the covariance function for the
GEV scale and shape parameters are updated similarly.\medskip

\noindent
\textbf{Step 6: Updating the projected Gaussian parameters}\\
To update the parameter $\tau_\theta$, we generate a proposal
$\tau_{\theta,p}$ from a log-normal distribution and compute the
acceptance probability
\begin{equation*}
  \alpha\{\tau_{\theta}, \tau_{\theta,p}\} = \min \left\{1, 
    \frac{\pi(\mathbf{R}, \boldsymbol{\theta} \mid \boldsymbol{\mu},
      \boldsymbol{\sigma}, \boldsymbol{\xi},
      \boldsymbol{\beta}_{\theta}, \tau_{\theta,p}, \rho,
      \lambda_{\theta}) \pi(\tau_{\theta, p} \mid
      \kappa_{\tau_\theta}^*,
      \theta_{\lambda_\theta}^*)}{\pi(\mathbf{R}, \boldsymbol{\theta}
      \mid \boldsymbol{\mu}, \boldsymbol{\sigma}, \boldsymbol{\xi},
      \boldsymbol{\beta}_{\theta}, \tau_{\theta, t}, \rho,
      \lambda_{\theta}) \pi(\tau_{\theta, t} \mid
      \kappa_{\tau_\theta}^*, \theta_{\lambda_\theta}^*)}
    \frac{\tau_{\theta, p}}{\tau_{\theta}}\right\}, 
\end{equation*}
a ratio of projected Gaussian likelihood times the ratio of the prior
densities with a correction due to the use of non symmetric proposal
distribution and where $\kappa_{\tau_\theta}^*$ and
$\theta_{\lambda_\theta}^*$ are respectively the shape and scale
parameters of the Gamma prior distribution. With probability
$\alpha\{\tau_{\theta}, \tau_{\theta, p}\}$, the $\tau_\theta$
component of $\boldsymbol{\psi}_{t+1}$ is set to $\tau_{\theta, p}$;
otherwise it remains at $\tau_{\theta, t}$. The parameter $\lambda$ is
updated similarly as well as the parameter $\rho$ except that, for the
latter parameter, we use the following symmetric proposal distribution
$\rho_p \sim \rho + U(-\epsilon_\rho, \epsilon_\rho)$ and consequently
no correction like $\rho_p / \rho$ is required.\medskip

\noindent
\textbf{Step 7: Updating the range parameters of the covariance
  function}\\
To update the parameter $\lambda_\mu$, we generate a proposal
$\lambda_{\mu,p}$ from a log-normal distribution and compute the
acceptance probability
\begin{equation*}
  \alpha(\lambda_{\mu}, \lambda_{\mu, p}) = \min
  \left\{1, \frac{\pi(\boldsymbol{\mu} \mid  \tau_{\mu},
      \lambda_{\mu, p},
      \boldsymbol{\beta}_{\mu})}{\pi(\boldsymbol{\mu} \mid
      \tau_{\mu}, \lambda_{\mu}, \boldsymbol{\beta}_{\mu})}
    \left(\frac{\lambda_{\mu, 
          p}}{\lambda_{\mu}} \right)^{k_{\lambda_\mu}^* - 1} \exp
    \left(\frac{\lambda_{\mu} - \lambda_{\mu,
          p}}{\theta_{\lambda_\mu}^*} \right) \frac{\lambda_{\mu,p}}{\lambda_{\mu}}\right\},
\end{equation*}
a ratio of multivariate Normal densities times the ratio of the prior
densities and that of the proposal densities and where
$\kappa_{\lambda_\mu}^*$ and $\theta_{\lambda_\mu}^*$ are respectively
the shape and the scale parameters of the Gamma prior distribution.
With probability $\alpha(\lambda_{\mu}, \lambda_{\mu,p})$, the
$\lambda_\eta$ component of $\boldsymbol{\psi}_{t+1}$ is set to
$\lambda_{\mu,p}$; otherwise it remains at $\lambda_{\mu}$. The range
parameters related to the scale and shape GEV parameters are updated
similarly. If the covariance family has a shape parameter like the
powered exponential or the Whittle--Mat\'ern covariance functions,
this is updated in the same way.

\bibliographystyle{apalike}
\bibliography{main}

\end{document}